\documentclass[%
 reprint,
longbibliography,
 amsmath,amssymb,
 aps,
prl,
twocolumn,
superscriptaddress
]{revtex4-1}

\usepackage[pdftex]{graphicx}
\usepackage{subfigure}
\usepackage{epstopdf}
\usepackage{graphicx}
\usepackage{dcolumn}
\usepackage{bm}
\usepackage[colorlinks,
linkcolor=blue, citecolor=blue, anchorcolor=blue,
urlcolor=blue,
]{hyperref}
\usepackage{braket}
\usepackage{color}
\usepackage{tikz}
\usepackage{textcomp}

\newcommand{\Yb}{$^{171}\textrm{Yb}^+~$}
\newcommand{\Ba}{$^{138}\textrm{Ba}^+~$}
\newcommand{\KDD}{$\textrm{KDD}_\mathrm{xy}~$}

\newcommand{\Unit}[1]{\,\mathrm{#1}}

\begin{document}

\title{Single ion-qubit exceeding one hour coherence time}

\author{Pengfei Wang}
\affiliation{Institute for Interdisciplinary Information Sciences, Tsinghua University, BeiJing 100084, China}
\author{Chun-Yang Luan}
\affiliation{Institute for Interdisciplinary Information Sciences, Tsinghua University, BeiJing 100084, China}
\author{Mu Qiao}
\affiliation{Institute for Interdisciplinary Information Sciences, Tsinghua University, BeiJing 100084, China}
\author{Mark Um}
\affiliation{Institute for Interdisciplinary Information Sciences, Tsinghua University, BeiJing 100084, China}
\author{Junhua Zhang}
\affiliation{Shenzhen Institute for Quantum Science and Engineering, and Department of Physics, Southern University of Science and Technology, Shenzhen 518055, P. R. China}
\affiliation{Institute for Interdisciplinary Information Sciences, Tsinghua University, BeiJing 100084, China}
\author{Ye Wang}
\affiliation{Institute for Interdisciplinary Information Sciences, Tsinghua University, BeiJing 100084, China}
\affiliation{Fitzpatrick Institute for Photonics, Electrical and Computer Engineering Department, Duke University, Durham, NC 27708, USA}
\author{Xiao Yuan}
\affiliation{Stanford Institute for Theoretical Physics, Stanford University, Stanford California 94305, USA}
\author{Mile Gu}
\affiliation{Centre for Quantum Technologies, National University of Singapore, Singapore 117543, Singapore}
\affiliation{School of Mathematical and Physical Sciences, Nanyang Technological University, Singapore 637371, Singapore}
\affiliation{Complexity Institute, Nanyang Technological University, Singapore 637335, Singapore}
\author{Jingning Zhang}
\affiliation{Beijing Academy of Quantum Information Sciences, Beijing 100193, China}
\author{Kihwan Kim}
\affiliation{Institute for Interdisciplinary Information Sciences, Tsinghua University, BeiJing 100084, China}

\date{\today}

\begin{abstract}
Realizing a long coherence time quantum memory is a major challenge of current quantum technology. Here, we report a single \Yb ion-qubit memory with over one hour coherence time, an order of improvement compared to the state-of-the-art record. The long coherence time memory is realized by addressing various technical challenges such as ambient magnetic-field noise, phase noise and leakage of the microwave oscillator. Moreover, systematically study the decoherence process of our quantum memory by quantum process tomography, which enables to apply the strict criteria of quantum coherence, relative entropy of coherence. We also benchmark our quantum memory by its ability in preserving quantum information, i.e., the robustness of quantum memory, which clearly shows that over 6000 s, our quantum memory preserves non-classical quantum information. Our results verify the stability of the quantum memory in hours level and indicate its versatile applicability in various scenarios.
\end{abstract}

\maketitle


Quantum coherence is a vital component for scalable quantum computation \cite{Lidar1998,DiVincenzo2001,Ladd10}, quantum metrology \cite{Kotler11,Degen2017}, and quantum communication  \cite{Briegel1998,Hartmann2007,Razavi2009,DuanMonroe12,Dudin2013}.
In practice, decoherence, loss of coherence in the computational basis, in the quantum system comes from the coupling with the surrounding environment and fluctuations of control parameters in quantum operations, which can lead to infidelity of quantum information processing, low sensitivity of quantum sensors, and inefficiency of quantum repeater based protocols in quantum communication networks. Limited coherence time may also undermine quantum-information applications such as quantum money~\cite{Wiesner83,Pastawski12}. It is thus of practical importance to have a stable quantum memory with a long coherence time.

Numerous experimental attempts have been made to enhance the coherence time of a quantum memory in a variety of quantum systems. With ensembles of nuclear spins in a solid, coherence time of up to order of an hour at room temperature \cite{Steger12,Saeedi13} and a few hours at liquid helium temperatures \cite{Zhong2015} have been reported. In a trapped ion system, cloud of trapped ions \cite{Bollinger91,Fisk95} demonstrated the coherence time of around 10 min. For a single qubit quantum memory, which is the essential building-block for quantum computers \cite{Cirac1995,Wright2019} and quantum repeaters \cite{Sangouard2009,Santra2019}, records of coherence time have been reported to the time scale of a minute in trapped ion-qubit \cite{Langer2005,Haffner05, harty2014high,kotler2014measurement}. The main limitation comes from the problem of qubit-detection inefficiency \cite{Epstein07,Wesenberg07,kotler2014measurement} due to motional heating of qubit-ions without Doppler laser-cooling. The limitation was addressed by sympathetic cooling by other species of ion, which allowed further improvements of coherence time to over 10 min with the support of dynamical decoupling \cite{YeWang17}. While there is no theoretical limit of the coherence time of quantum systems, however, it remains a major technological challenge to further enhance the quality of a trapped ion quantum memory.

Here we address this challenge by improving the coherence time of a \Yb ion-qubit memory from 10 min to over one hour. This is achieved by identifying and suppressing the three dominant error sources: magnetic-field fluctuation, phase noise of the local oscillator, and microwave leakage for qubit operation. Furthermore, with the capability of full control on single qubit, we systematically study the decoherence process of the quantum memory by quantum process tomography. Typically, the decoherence process has been characterized by the coherence time $T_2$ at which the Ramsey contrast, corresponding to the size of the off-diagonal entry in the qubit density-matrix, decays to 1/e \cite{Steger12,Saeedi13,Zhong2015,Bollinger91,Fisk95,YeWang17}. We experimentally study the decoherence dynamics by relevant quantum channels of depolarization and dephasing, which allows us to use recently developed coherence quantifiers \cite{YuanXiao2015,Winter2016,Streltsov2017}. We also apply the criteria of the robustness of quantum memory that quantifies how well a memory preserves quantum information \cite{XiaoYuan19}, demonstrating that entanglement can still remain after 6000 seconds.

\begin{figure}[htp]
\centering
\subfigure[]{\includegraphics[width=0.45\textwidth]{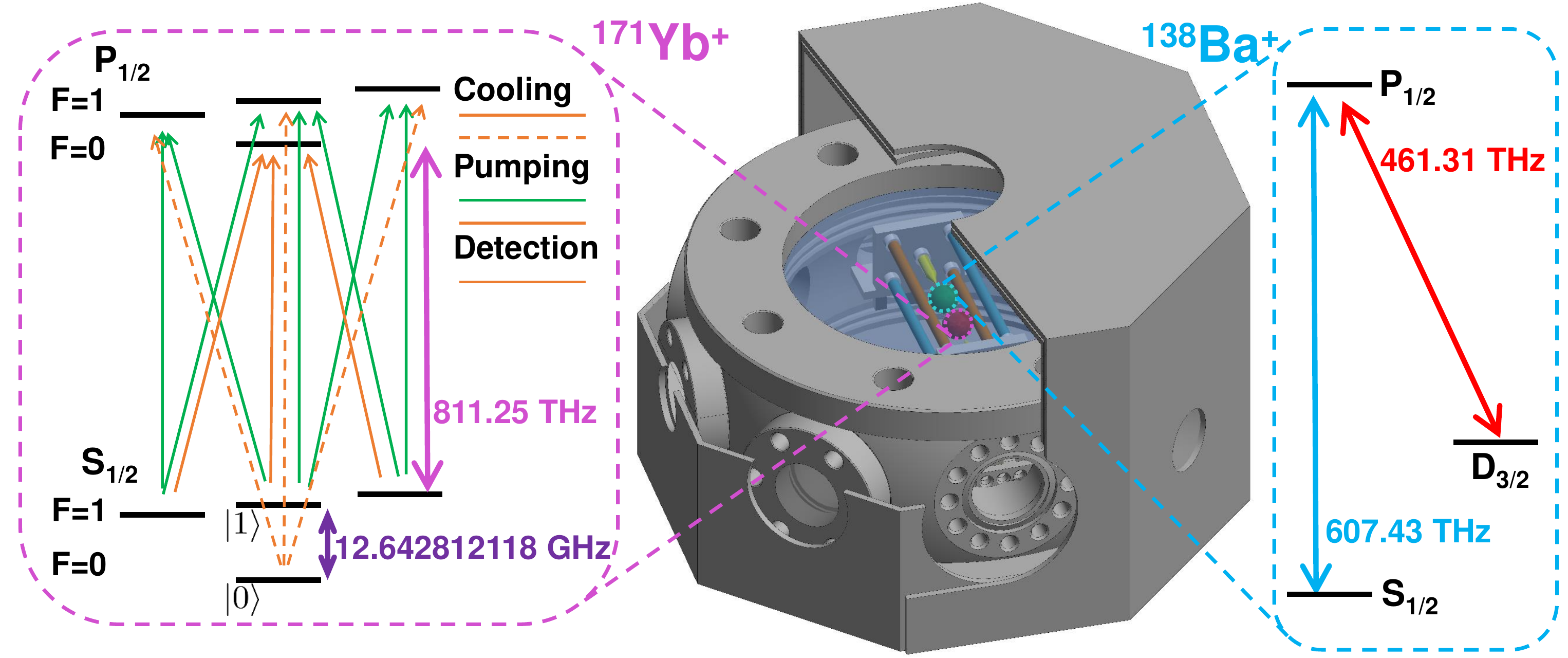}}\\
\subfigure[]{\includegraphics[width=0.45\textwidth]{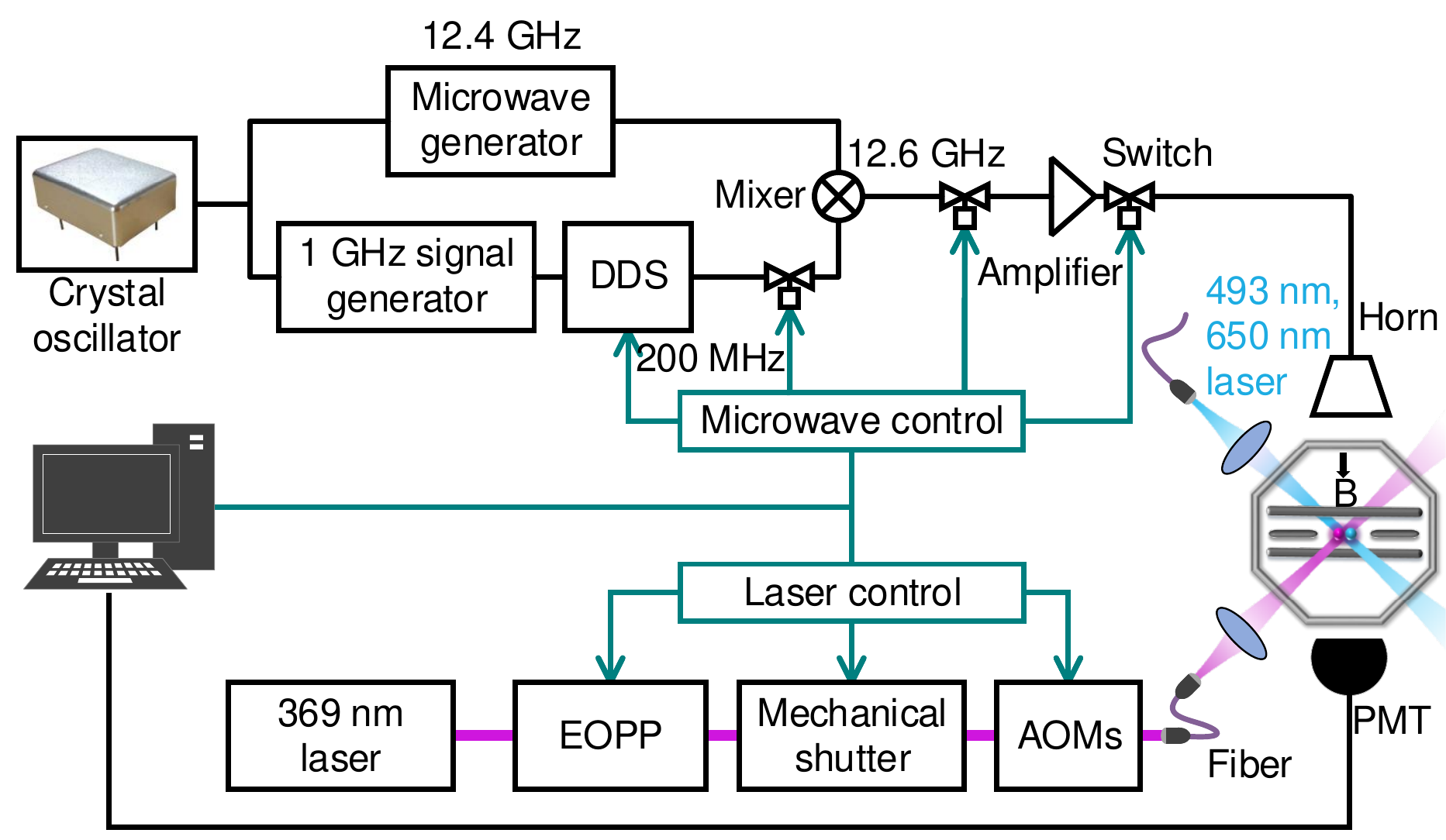}}
\caption{Experimental setup.
(a), Energy levels of \Yb and \Ba ion and cutaway view of the \textmu-metal shielding enclosing octagon chamber. The shielding has ten holes, where two holes for the connection of vacuum pump and helical resonator and the other eight holes with diameter from 20 mm to 40 mm for the access of laser beams, microwave and imaging system.
(b), Schematic diagram for the control of microwave and laser beams. We use a crystal oscillator (SIMAKE SMK3627OCHFM OCXO) to reference the microwave generator and Direct Digital Synthesize (DDS) through 1 GHz signal generator. The microwave of 12.6 GHz is generated by mixing 200 MHz signal from DDS board and 12.4 GHz from microwave generator, which is amplified and applied to ions through a horn. All three microwave switches are used to reduce microwave leakage. For 369 nm laser beams, we use (acousto-optic modulators) AOMs to generate basic operate lasers. We use Electro-Optic pulse picker (EOPP), mechanical shutter and single-mode fiber to reduce laser leakage. Magnetic field direction is in the radial direction. We detect the qubit state with a photomultiplier tube (PMT).}
\label{shielding}
\end{figure}

In our experiment, we load one \Yb ion and one \Ba ion in a four-rod Paul trap as shown in Fig.\ref{shielding}(a). Two hyperfine levels of the \Yb ion in the $\rm{S}_{1/2}$  manifold are used to encode the qubit with $\{\ket{0}\equiv\ket{F=0, m_F=0},\ket{1}\equiv\ket{F=1, m_F=0}\}$ and a frequency difference of $12642812118 + 310.8 B^2\Unit{Hz}$, where $B$ is the magnetic field in Gauss. As a sympathetic cooling ion, \Ba is used since it has a similar atomic mass with {$^{171}\textrm{Yb}^+$}, which can be used to efficient cooling. We apply Doppler-cooling laser beams on the \Ba ion all the time, which provides continuous cooling for the whole system. In this way, we can measure the final state of the \Yb qubit by standard fluorescence detection technique without losing any detection fidelity \cite{Epstein07,Wesenberg07,kotler2014measurement}.

We suppress the ambient noise of the magnetic field by installing a magnetic-field shielding with a permanent magnet \cite{ruster2016long}. We enclose our main vacuum chamber that contains the Paul trap with a two-layer of \textmu-metal shielding shown in Fig.\ref{shielding}(a). By using a fluxgate meter, we observe more than 40 dB attenuation at 50 Hz inside the shielding, which is the main frequency of noise in the lab due to AC power-line. To generate stable magnetic field of 5.8 G, we replace coils with a $\textrm{Sm}_2\textrm{Co}_{17}$ permanent magnet,  which has a temperature dependence of $-0.03$ \%/K \cite{ruster2016long}. The magnetic-field strength can be adjusted by changing the position of the magnet from the location of ions. After these modifications, we observe the coherence time of the field-sensitive Zeeman qubit is increased to be more than 30 ms. We study the noise spectrum by dynamical decoupling sequences \cite{Biercuk09,kotler2013nonlinear} and observe that noise of 50 Hz and 150 Hz are below 16 \textmu G and 32 \textmu G, respectively \cite{SM}.

We perform coherent manipulation of the qubit by applying a resonant microwave. Qubit coherence is typically measured by the contrast of Ramsey fringe, which requires control and interrogation of the system by a local oscillator that can bring in phase noise \cite{Ball2016}. In our case, this part of noise is determined by microwave signal generator and its reference. For microwave signal, phase noise in low frequency regime is mainly determined by those of the reference signal \cite{Agilent}. We use a crystal oscillator as the reference, which has one order smaller than our previous oscillator of Rb clock in Allan deviation at 1 s observation time \cite{YeWang17}.

We also find that leakage of the laser and microwave can introduce relaxation of the qubit memory. In order to suppress the leakage of \Yb-ion resonant laser beams, we use acousto-optic modulators, electro-optic pulse picker and a mechanical shutter \cite{YeWang17}. For microwave control, we suppress the microwave output by 164 dB after turning off all the switches. With $\pi$ pulse duration of 175 \textmu s, the effect of leakage is negligible for 0.4 s  pulse interval time, which would be further suppressed by dynamical decoupling pulses. 

\begin{figure}[h]
\includegraphics[width=0.45\textwidth]{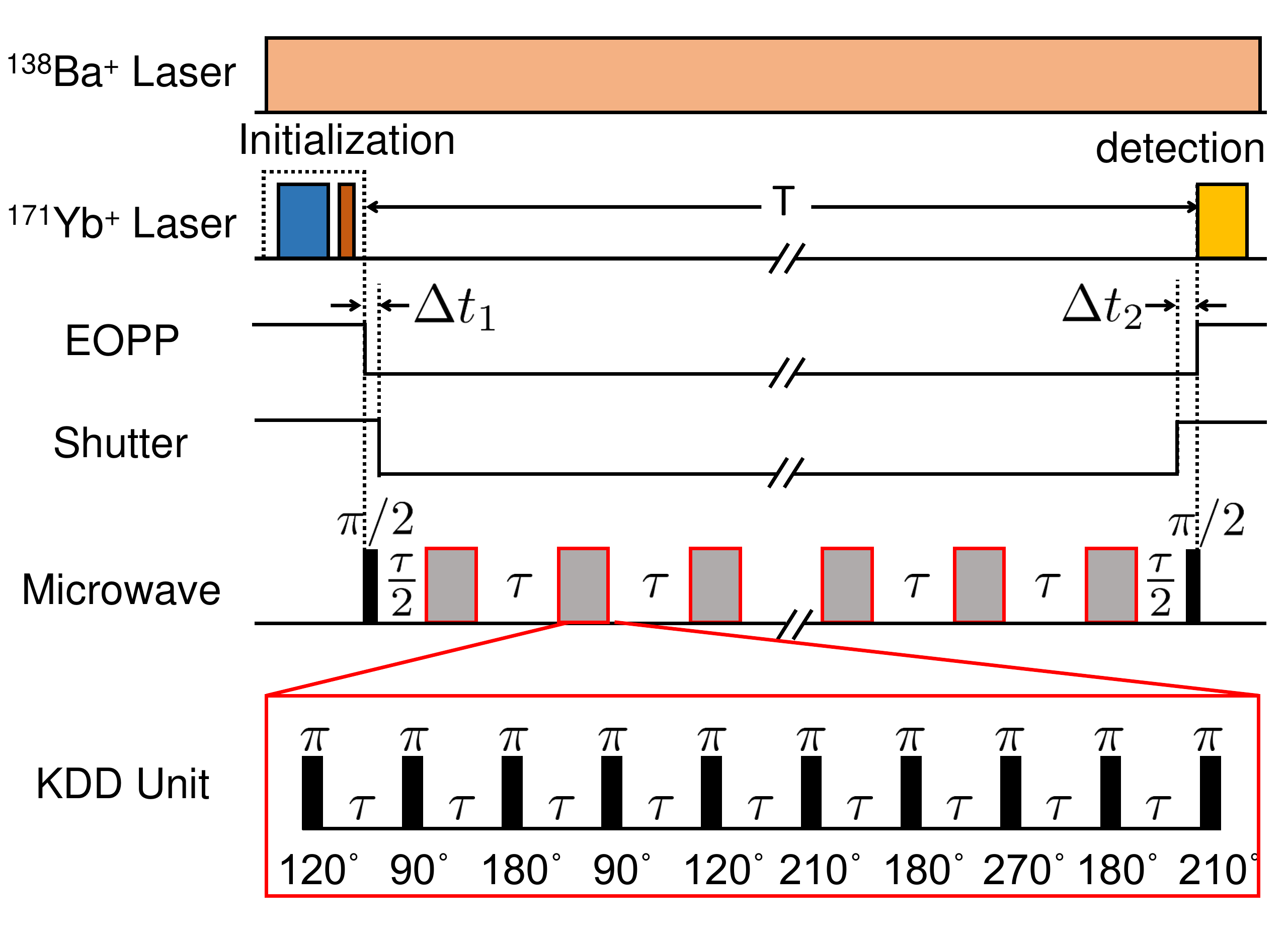}
\caption{Experimental sequence. Cooling laser beams for \Ba ion are applied during the whole sequence. For \Yb ion, we first initialize the qubit and then start to apply the microwave pulses. All the \KDD pulses are inserted between two $\pi/2$ pulses of Ramsey sequence. Blue and brown blocks represent Doppler cooling and optical pumping pulses for \Yb ion. EOPP and shutter are closed after state initialization and opened before state readout, where the time delays between them are shown as $\Delta t \approx 10$ ms, which is mainly caused by limited speed of mechanical shutter. Gray blocks represent \KDD units. T is the total measurement time, and $\tau$ is the inter-pulse delay. Each \KDD unit has ten $\pi$ pulses, where the first and the second five pulses represent $\sigma_{\rm x}$- and $\sigma_{\rm y}$-rotation, respectively. Therefore, the second five pulses have  $90^\circ$ phase shift from the first five. We choose total number of \KDD units even to make sure all the \KDD pulses are identity operation in the ideal case. In the end, we use detection laser to measure the qubit state. }
\label{sequence}
\end{figure}

We measure the coherence time of the \Yb ion-qubit by observing the dependence of Ramsey contrasts on the storage time. The experimental sequence is shown in Fig.~\ref{sequence}. As discussed above, cooling laser beams for \Ba are applied during the whole sequence. We initialize the state of the \Yb ion-qubit to $\ket{0}$ by the standard optical pumping technique, apply the $\pi/2$-Ramsey pulses, and detect the probability in $\ket{1}$ state by the standard state-dependent fluorescence method.  We note that we have a detection efficiency of 98.6 $\%$, which is corrected by the calibrated error magnitude with the uncorrelated error assumption as shown in Ref.~\cite{shen2012correcting}.

To enhance coherence time, we apply the dynamical decoupling scheme  \cite{Saeedi13,Zhong2015,Khodjasteh13,Biercuk09,kotler2013nonlinear,Souza11,haeberlen1976high,YeWang17}, which is based on spin-echo that uses single $\pi$ pulse to compensate low frequency noise. Performance of dynamic decoupling pulses is described by the filter function $\widetilde{y}(\omega,T)=\frac{1}{\omega}\sum _{j=0}^{T/\tau} (-1)^j \left(\mathrm{e}^{i\omega t_{j}}-\mathrm{e}^{i\omega t_{j+1}}\right)$,
with  $t_0=0$, $t_{T/\tau+1}=T$, $t_j=(j-0.5)\tau$ when $1\le j\le T/\tau$, and $\tau$ is the interval of pulses. Then Ramsey fringe contrast \cite{Biercuk09} is $W(T)= \mathrm{e}^{-\frac{2}{\pi}\int _o^{\infty }S(\omega){\left|\widetilde{y}(\omega,T)\right|}^2 d\omega}$ with $S(\omega)$ being the noise spectrum density.
In our experiment, we use \KDD (Knill dynamical decoupling) \cite{Zhong2015,YeWang17,Souza11} pulses, where all the pulses are equally spaced and have periodic phases as shown in Fig.~\ref{sequence}. Filter function of the \KDD\ pulses has a peak at the frequency $\omega=\frac{\pi}{\tau}$. Most part of the  noise is suppressed except the part with frequencies around the peak, which is instead amplified. When the total time T is fixed, the position of the peak is determined by the pulse interval, which can be optimized depending on the noise spectrum. After comparing different parameters, we choose 0.4 s as the pulse interval, which leads the peak of the filter function at $2\pi \times 1.25$ Hz.

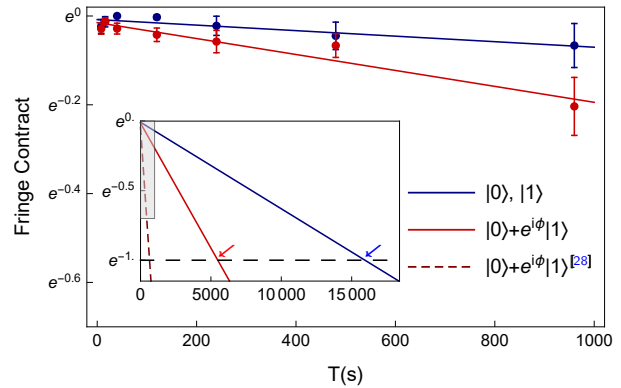
\begin{figure}[ht]
\begin{tikzpicture}
\tikzstyle{every node}=[font=\tiny]
    \draw (0, 0) node[inner sep=0] {\includegraphics[width=0.45\textwidth]{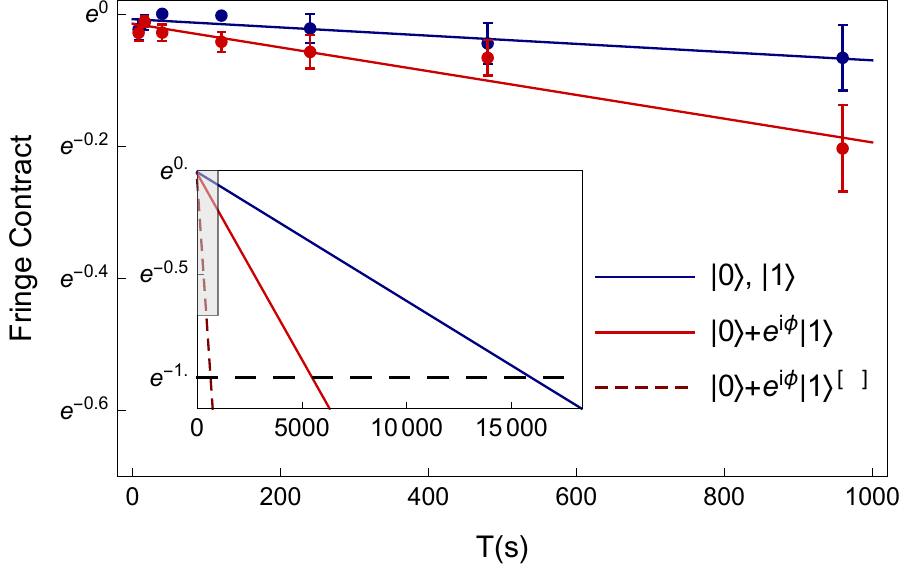}};
    \draw (3.58, -0.85) node {\textsf{\cite{YeWang17}}};
    \draw (-1.12, -0.7) node {{\color{red}$\pmb{\swarrow}$}};
    \draw (0.83, -0.7) node {{\color{blue}$\pmb{\swarrow}$}};
\end{tikzpicture}
\caption{The evolution of Ramsey fringe contrast with \KDD pulse sequences. Blue points are from the initial states of $\ket{0}$ and $\ket{1}$, and red points are from $\ket{0}+\ket{1}$, $\ket{0}+i\ket{1}$, $\ket{0}-\ket{1}$, and $\ket{0}-i\ket{1}$, where $\phi = 0$, $\frac{\pi}{2}$, $\pi$, and $\frac{3\pi}{2}$, respectively. Error bars are standard deviations. Each initial state at each data point repeats 30 to 100 times. The solid lines are the fitting results by exponential decay function. Inset shows fitting results in a longer time range. Rectangle shadow indicates the enlarged area in the big figure. The red-dashed line indicates the previous result of superposition states \cite{YeWang17}. The black-dashed line indicates the $1/\rm{e}$ threshold. The red and blue arrows indicate times when threshold are reached.}
\label{raw data}
\end{figure}

With different initial states, we show the time dependence of the Ramsey contrast up to 16 min in Fig.~\ref{raw data}. By assuming an exponential decay of the Ramsey contrast, we find the  coherence time of states  $\ket{0}$ and $\ket{1}$ to be 15943 $\pm $ $3189.5$ s. Other four superposition states ($\phi = 0$, $\frac{\pi}{2}$, $\pi$, and $\frac{3\pi}{2}$ shown in the legends of Fig.~\ref{raw data} ) have a coherence time of 5487 $\pm$ $667.6$ s. Both of the uncertainties are from fitting errors. As show in the inset of Fig.~\ref{raw data}, the coherence time is increased by an order of magnitude compared to the previous state-of-the-art result~\cite{YeWang17}.

\begin{figure}[ht]
\subfigure[]{
\begin{tikzpicture}
\tikzstyle{every node}=[font=\footnotesize]
    \draw (0, 0) node[inner sep=0] {\includegraphics[width=0.45\textwidth]{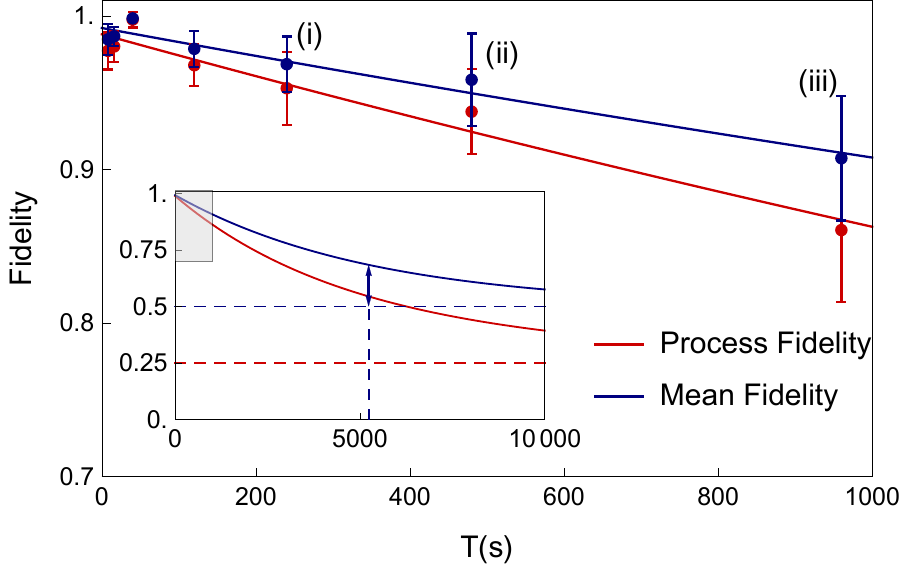}};
    \draw (-1.0, -0.4) node {\textsf{{\color{blue}{Final mean fidelity}}}};
    \draw (-1.0, -0.9) node {\textsf{{\color{red}{Final process fidelity}}}};
    \draw (-0.95, 0) node {\color[rgb]{0.25,0.25,0.6}\textsf{\tiny{$\bm{\frac{1}{\rm{e}}}$}}};
\end{tikzpicture}
}\\
\vspace{-2 pt}

\subfigure{
\begin{tikzpicture}
\tikzstyle{every node}=[font=\footnotesize]
    \draw (0.05, 0) node[inner sep=0] {\includegraphics[width=0.44\textwidth]{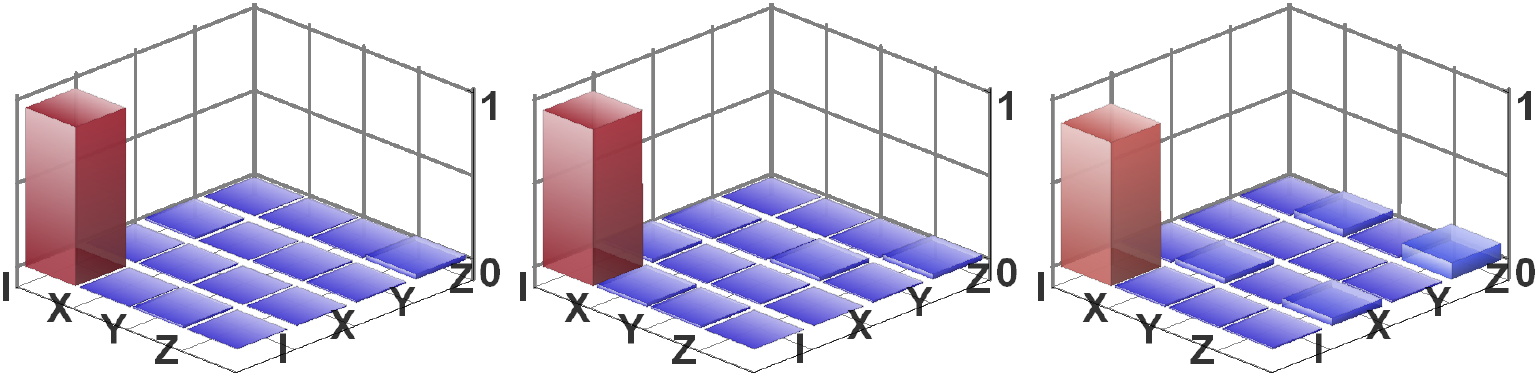}};
    \draw (-2.6, 1.2) node {\textsf{(i)}};
    \draw (0.05, 1.2) node {\textsf{(ii)}};
    \draw (2.7,1.2) node {\textsf{(iii)}};
    \draw (-4.2, 0) node {\textsf{(b)}};
\end{tikzpicture}}\\
\vspace{-10 pt}
\subfigure{
\begin{tikzpicture}
\tikzstyle{every node}=[font=\footnotesize]
    \draw (0.05, 0) node[inner sep=0] {\includegraphics[width=0.44\textwidth]{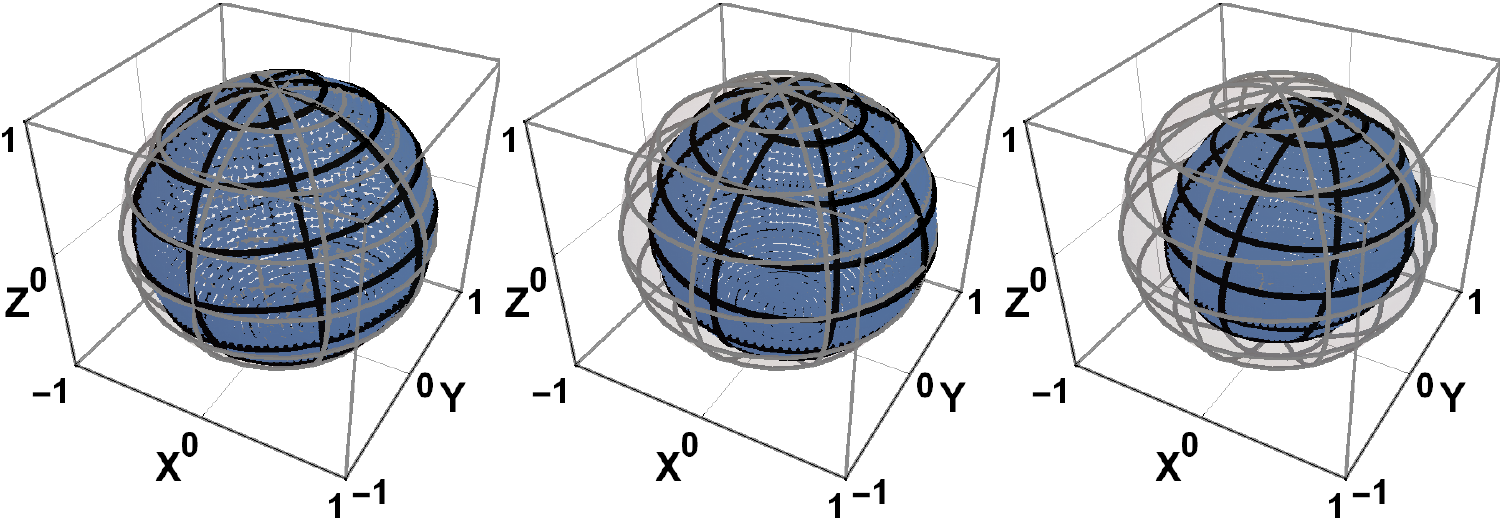}};
    \draw (-4.2, 0) node {\textsf{(c)}};
\end{tikzpicture}}\\

\caption{The evolution of process and mean fidelities and examples of quantum process tomography results. (a), Red and blue points represent process and mean fidelities, respectively. The red line is the fitting result of Eq. (\ref{eq:quantumchannel}). The blue line is the fitting result of exponential decay function. Inset shows fitting results in a longer time range. Rectangle shadow indicates the enlarged area in the big figure. The red and blue dashed horizontal lines indicate the process fidelity and mean fidelity of the final state, where the system lost all the quantum information. The blue vertical lines indicate the time point when mean fidelity decays to $1/\rm{e}$ threshold. (b), Real part of the process matrix after a storage time of (i): 4 min, (ii): 8 min and (iii): 16 min. The largest diagonal element of the process matrix is the identity operation part, $\chi^\textrm{exp}_{11}$, which is the  process fidelity $F_\mathrm{p}$. (c), State evolution represented in the Bloch sphere after a storage time of (i): 4 min, (ii): 8 min and (iii): 16 min. Gray meshed spheres represent the initial state and blue spheres represent the output states after corresponding storage time.}
\label{tomography}
\end{figure}
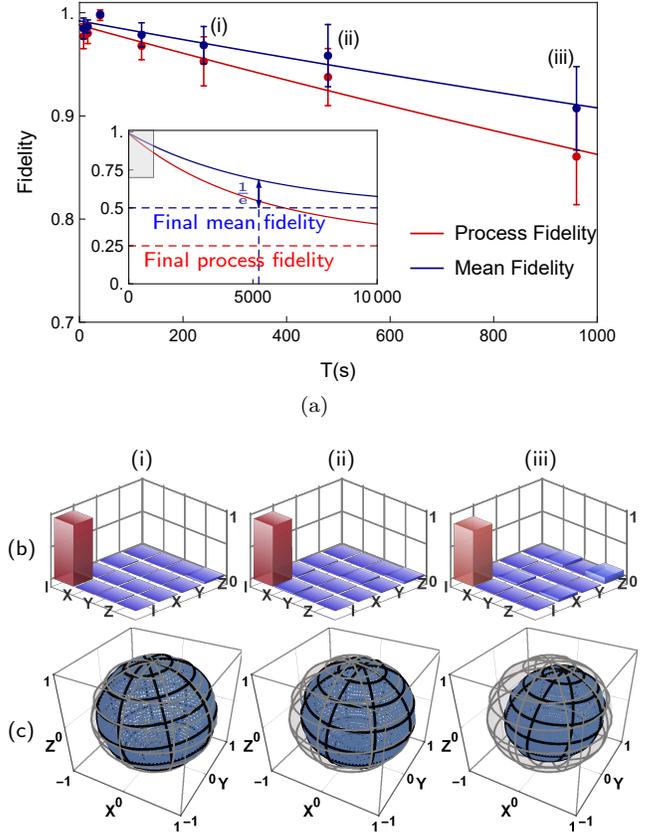

We further analyze the decoherence process by performing quantum process tomography at different storage time following the Refs.  \cite{Fiur01,nielsen2010quantum}. For a quantum process $\varepsilon$, we consider its process $\chi$ matrix, which is defined by $\varepsilon (\rho)=\sum_{mn}\chi_{mn} \hat{E}_m \rho \hat{E}_n^\dagger$ with $\hat{E}_m\in \{\hat{I}, \hat{X}, \hat{Y}, \hat{Z}\}$ \cite{nielsen2010quantum} (see also Supplemental Material~\cite{SM}).
We observe the time dependence of the process matrix as shown in Fig.~\ref{tomography}(a). The ideal process of quantum memory is described by $\chi^{\rm id}_{mn}=\delta_{m,1}\delta_{n,1}$. With the experimentally measured process matrix $\chi^\textrm{exp}$, we can obtain the process fidelity $F_\mathrm{p} =\mathrm{Tr}(\chi^\textrm {id}\chi^\textrm{exp})= \chi^\textrm{exp}_{11}$. The infidelity mainly comes from the dephasing and depolarization effects. The process with these two noises can be described by the following matrix as
\begin{equation}
\left[
\begin{array}{cccc}
   \frac{1+2\mathrm{e}^{-t/T_2}+\mathrm{e}^{-t/T_1}}{4}  &  0  &  0  &  0  \\
     0 &  \frac{1-\mathrm{e}^{-t/T_1}}{4}  &  0  &  0  \\
     0 &  0  &  \frac{1-\mathrm{e}^{-t/T_1}}{4}   &  0  \\
     0 &  0  &  0  &  \frac{1-2\mathrm{e}^{-t/T_2}+\mathrm{e}^{-t/T_1}}{4}  \\
\end{array}
\right],\label{eq:quantumchannel}
\end{equation}
where $T_1$ and $T_2$ are depolarizing and dephasing time, respectively  \cite{Sarvepalli2009}. The process matrix describes the quantum memory that in the beginning, no decoherence, and $t\gg T_1, T_2$ any initial states are changed to fully mixed state. By fitting  the experimental process tomography results with the above process matrix of Eq. (\ref{eq:quantumchannel}), we obtain $T_1$ = 11902 $\pm$ 2209.3 s and $T_2$ = 4235 $\pm$ 575.7 s (See more details in Supplemental Material~\cite{SM}).
We also plot the model of Eq. (\ref{eq:quantumchannel}) and the experimental data in Fig.~\ref{tomography}(a).

Another way to benchmark a quantum memory is via the mean fidelity,
$F_\textrm{mean}=\langle \mathrm{Tr}(\rho\varepsilon (\rho ))\rangle_{\rho}$,
which is the averaged output fidelity with all possible input states $\rho$~\cite{Gilchrist2005Distance, Brien2004Quantum,Riebe2006Process}.
We use the Monte Carlo method to get the mean fidelity with $10^5$ different input states, generated by uniformly sampled random unitary operations according to the Haar measure~\cite{Maris2009}. As shown in Fig.~\ref{tomography}(a), we obtain the coherence time, the time constant of fitted exponential decay-function, 5235 $\pm$ 495.5~s for the mean fidelities. We can also obtain $F_\textrm{mean}$ by using the relation between the process fidelity and the mean fidelity as $F_\textrm{mean}=(d F_\textrm{p}+1)/(d+1)$ with $d=2$ for qubits \cite{Gilchrist2005Distance}. When we use the formula, We observe a little longer time of decaying to 1/e as 5635 $\pm$ 652.8~s.

\begin{figure}[ht]
\begin{tikzpicture}
\tikzstyle{every node}=[font=\footnotesize]
    \draw (0, 0) node[inner sep=0] {\includegraphics[width=0.45\textwidth]{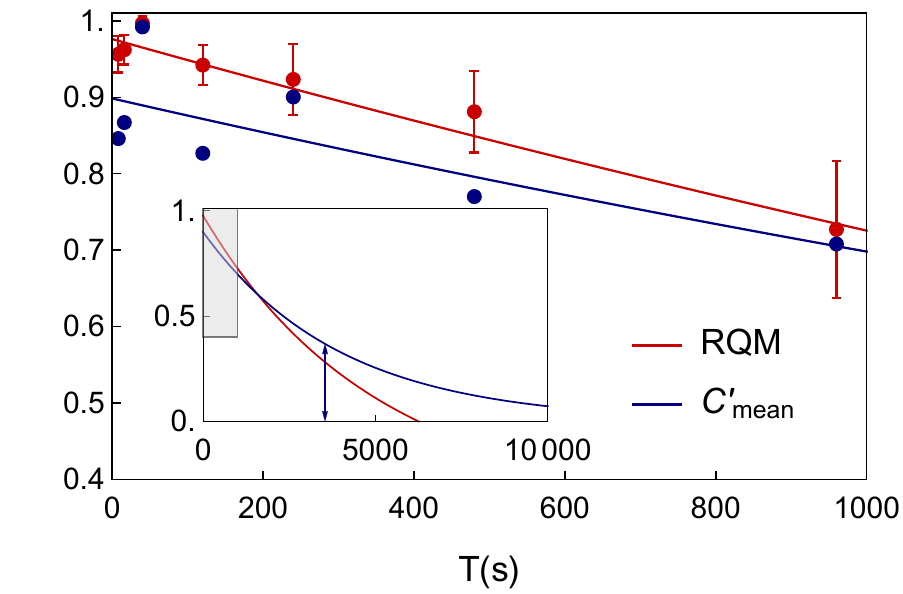}};
    \draw (-1.4, -0.8) node {\color[rgb]{0.25,0.25,0.6}\textsf{\small{$\bm{\frac{1}{\rm{e}}}$}}};
\end{tikzpicture}
\caption{The evolution of robustness of quantum memory and mean REC ratio. Red and blue points are robustness of quantum memory and mean ratio of REC, respectively. Red line is the calculated RQM from the process fidelity $F_{\rm p} and $blue line is exponential fitting result. Inset shows fitting results in a longer time range. Rectangle shadow indicates the enlarged area in the big figure. The blue vertical lines indicate the time point when mean REC ratio decays to $1/\rm{e}$ threshold.}
\label{Robustness}
\end{figure}

We also measure the robustness of quantum memory (RQM)~\cite{XiaoYuan19}, which quantifies how well the memory preserves quantum information. The RQM of $N$ is defined as
$R(N)= \mathop{\min}_{M\in \rm EB}\left\{s\ge0\bigg|\frac{N+sM}{s+1}\in {\rm EB}\right\}$, with EB being the set of entanglement-breaking (or equivalently measure-and-prepare) memories that  deterministically break entanglement. We note that EB memories cannot maintain quantum information and hence have zero RQM. We calculate the RQM of our system based on the process matrix. In our experiments, off-diagonal elements in the process matrix is negligible, which simplifies the RQM to $\mathop{\max}\{2 F_{\rm p}-1,0\}$. As shown in Fig.~\ref{Robustness}, the RQM of our system lasts 6256~s before  it decays to zero by exponential fitting.

Finally, we measure how our quantum memory can preserve quantum coherence~\cite{YuanXiao2015,Winter2016,Streltsov2017}. We focus on the relative entropy of coherence (REC)~\cite{Winter2016}, $C(\rho)=S(\Delta (\rho))-S(\rho)$, with $\Delta(\rho)=\sum _{i}\bra{i}\rho\ket{i}\ket{i}\bra{i}$, $\{\ket{i}\}$ being the computational basis, and $S(\rho)=-\mathrm{Tr}(\rho \log_2\rho)$ being the Von Neumann entropy. Based on the process matrix $\chi^\textrm{exp}$, We numerically calculate the REC ratio between output state and the input state. We consider the mean REC ratio $C'_\textrm{mean}=\langle C({\varepsilon (\rho )})/C(\rho)\rangle_{\rho}$ by averaging over $10^5$ random input states. Note that we only consider states with REC larger than 0.01.
As shown in Fig.~\ref{Robustness}, the mean REC ratio decays to 1/e after 3539 $\pm$ 1141.9~s by exponential fitting.
The large fluctuation of the results mainly stems from stringent sensitivity of the REC to small errors in estimation of the process matrix. 

In this paper, we report a trapped-ion based single qubit quantum memory with over one hour coherence time, an order of magnitude enhancement compared to the state-of-the-art record~\cite{YeWang17}. Further enhancement may be achieved by improving the stability of the classical oscillator. Other technical limitations such as light scattering from the cooling laser beams of \Ba ion, magnetic field noise, microwave leakage and the imperfection of \KDD pulses are expected to set bounds of over ten hours \cite{SM} for our current experimental system. Ultimately, the frequency shift by ion-hopping and collisonal shift by background gases would be the limitation \cite{SM}, which can be further suppressed by locating the ion trap system in a cryostat of 4 K. For the general purpose of quantum memory, it is necessary to increase the number of ion-qubits. We find that hopping causes serious problems, but it can be also suppressed by the ion-trap in the cryostat. The micromotion should be carefully addressed for the multi-ion systems, which requires a more sophisticated trap to individually compensate the micromotions.\\

\section{acknowledgments}
We thank Roee Ozeri, Rene Gerritsma, Jianwei Zhang and Jizhe Han for helpful discussions. This work was supported by the National Key Research and Development Program of China under Grants No. 2016YFA0301900 and No. 2016YFA0301901, the National Natural Science Foundation of China Grants No. 11574002, No. 11974200, and No. 11504197, Singapore Ministry of Education through Tier 1 Grant No. RG190/17, the Singapore National Research Foundation through Fellowship No. NRF-NRFF2016-02, and NRF-ANR Grant No.
NRF2017-NRF-ANR004 VanQuTe. 

%

\section{Supplementary Information}

\section{Spontaneous emission, hopping and collision}

{\it Spontaneous emission of other lasers:}
the spontaneous emission rate can be estimated by \cite{ozeri2005hyperfine,uys2010decoherence,campbell2010ultrafast,YeWang17}:
\begin{equation}
\Gamma_{spon}=\frac{\gamma g^2}{6}(\frac{1}{\Delta^2_{\mathrm{D1}}}+\frac{2}{(\Delta_{\mathrm{FS}}-\Delta_{\mathrm{D1}})^2})
\end{equation}
where $\gamma\approx2\pi\times20$ MHz is the spontaneous emission rate from the $^2\rm{P}$ states, $g=\frac{\gamma}{2}\sqrt{I/(2I_{\mathrm{sat}})}$, $\Delta_{\mathrm{HF}}=12.6$ GHz, $\Delta_{\mathrm{FS}}= 2\pi\times 100$ THz. For 493 nm laser, power $P=35$ \textmu W, beam waist $\omega =31.4$ \textmu m, $I_{493}=21.8 I_{\mathrm{sat}}$, $\Delta_{\mathrm{D1}}=2\pi\times$ 203.8 THz, then we get a  scattering rate of $1.09\times 10^{-6}$ Hz. For the 650 nm laser, power $P=66$ \textmu W, beam waist $\omega =22.9$ \textmu m, $I_{650}=75.5 I_{\mathrm{sat}}$, $\Delta_{\mathrm{D1}}=2\pi\times 349.9$ THz, scattering rate $1.29\times 10^{-6}$ Hz. So 493 nm and 650 nm laser have negligible effect for hours level coherence time.

{\it Ion hopping:} Hopping of the ion happens about 10 min. Magnetic field difference between two ion positions is $60 \mu$G, which introduces the frequency difference of 36 mHz. The dynamical decoupling pulses with the interval of 0.4 s can compensate the frequency changes in about 10 minutes and we do not observe any limitation from hopping problem.

{\it Collision of background gas:} According to ref.~\cite{Hankin2019}, a $^{27}\textrm{Al}^+$ optical transition clock have a
mean frequency shift of order $10^{-16}$ caused by 38 nPa room-temperature H$_2$ background gas collision. Fractional frequency shift is always below $10^{-14}$ level with 10 s probe time. It will be smaller in our system due to sympathetic cooling, smaller  hyperfines energy difference and high vacuum of below 10 nPa. So this part shift is negligible for hour levele measurement time.

\section{Magnetic field noise}
\begin{figure}[ht]
\centering
\includegraphics[width=0.45\textwidth]{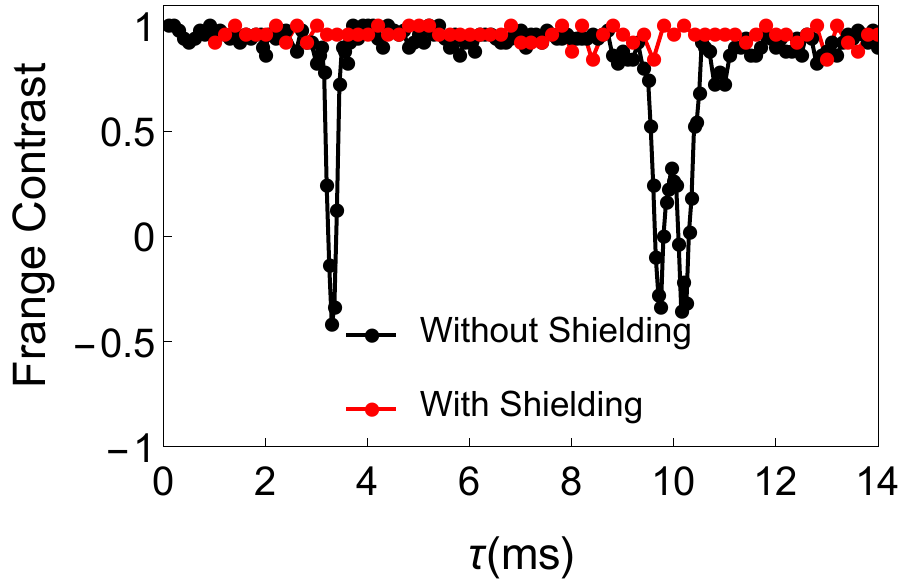}
\caption{Suppression of magnetic-field noise by magnetic-field shielding. We use 31 CPMG (Carr, Purcell, Meiboom and Gill)  \cite{haeberlen1976high} pulses to accumulate the AC magnetic-field noise. Black and red points represent without shielding \cite{YeWang17} and with shielding, respectively. Two deeps at 3.3 ms and 10 ms are caused by 150 Hz and 50 Hz noise, respectably. }
\label{noise}
\end{figure}

As shown in Fig.~\ref{noise}, we get a noise of 50 Hz and 150 Hz below 16 \textmu G and 32 \textmu G, respectively after installing magnetic field shielding and permanent magnet.

\section{Procedure of quantum process tomography of a single ion-qubit}
For our single ion-qubit memory, we measure the $\chi$ matrix by preparing four different input states $\ket{0}$, $\ket{1}$, $(\ket{0}+\ket{1})/\sqrt{2}$, $(\ket{0}+i\ket{1})/\sqrt{2}$, applying the memory, and finally measuring the output states with four measurements $I$, $X$, $Y$ and $Z$.

\section{Process matrix evolution}
\begin{figure}[ht]
\includegraphics[width=0.45\textwidth]{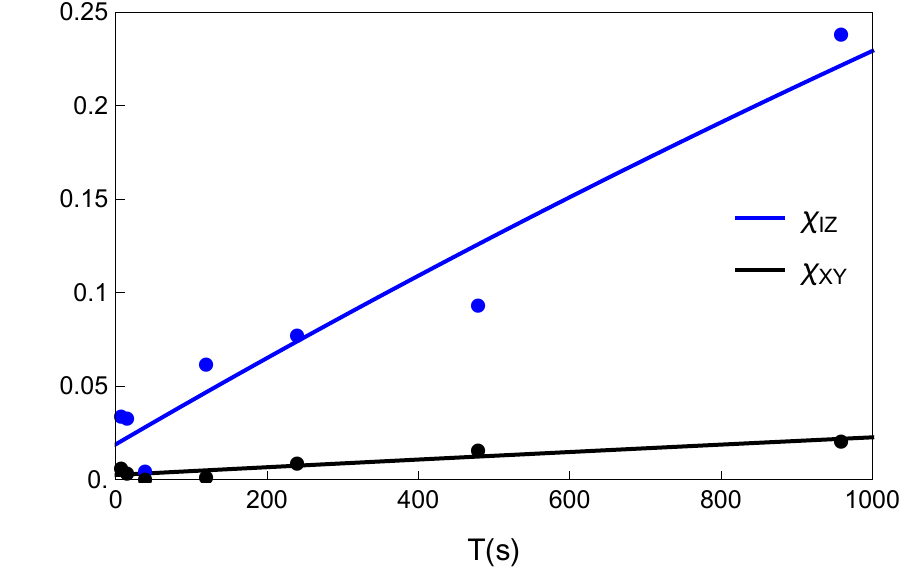}
\caption{Fitting of process matrix elements evolution. Here $\chi_{\textrm{IZ}}=1-(\chi_{11}-\chi_{44})$, and $\chi_{\textrm{XY}}=0.5(\chi_{22}+\chi_{33})$.}
\label{dephasing and depolization}
\end{figure}

As shown in Fig.~\ref{dephasing and depolization}, we fit the process matrix evolution with Eq. (\ref{eq:quantumchannel}) and obtain $T_2$ = 4235 $\pm$ 575.6 s and $T_1$ = 11902 $\pm$ 2209.3 s.

\end{document}